\documentclass[twocolumn,superscriptaddress,aps,preprintnumbers,amsmath,amssymb,prl,nofootinbib]{revtex4-2}

\usepackage{dcolumn}
\usepackage{graphicx}
\usepackage{epstopdf}
\usepackage{bm}
\usepackage{color}
\usepackage{mathrsfs}
\usepackage{physics}
\usepackage{comment}
\usepackage{here}

\def\Mpl{M_{\rm Pl}}
\def\GeV{{\rm GeV}}
\def\reh{{\rm reh}}
\def\rec{{\rm rec}}

\def\Mpc{{\rm Mpc}}
\def\mcA{\mathcal{A}}

\begin{document}

\title{
Low-Scale Inflationary Magnetogenesis
without Baryon Isocurvature Problem
}

\author{Kazuki Yanagihara}
\email{kazuki-y@ruri.waseda.jp}
\affiliation{Department of Physics, Waseda University, 3-4-1 Okubo, Shinjuku, Tokyo 169-8555, Japan}

\author{Fumio Uchida}
\email{fuchida@post.kek.jp}
\affiliation{Research Center for the Early Universe, The University of Tokyo, Bunkyo, Tokyo 113-0033, Japan}
\affiliation{Department of Physics, Graduate School of Science, The University of Tokyo, Bunkyo, Tokyo 113-0033, Japan}

\author{Tomohiro Fujita}
\email{tomofuji@aoni.waseda.jp}
\affiliation{Waseda Institute for Advanced Study, Waseda University, 1-6-1 Nishi-Waseda, Shinjuku, Tokyo 169-8050, Japan}
\affiliation{Research Center for the Early Universe, The University of Tokyo, Bunkyo, Tokyo 113-0033, Japan}

\author{Shinji Tsujikawa}
\email{tsujikawa@waseda.jp}
\affiliation{Department of Physics, Waseda University, 3-4-1 Okubo, Shinjuku, Tokyo 169-8555, Japan}

\begin{abstract}

\noindent
Primordial magnetogenesis is an intriguing possibility to explain the origin of 
intergalactic magnetic fields (IGMFs). However, the baryon isocurvature problem 
has recently been pointed out, ruling out all magnetogenesis models operating 
above the electroweak scale. In this letter, we show that lower-scale 
inflationary scenarios with a Chern-Simons coupling can evade this problem. 
We propose concrete inflationary models whose reheating temperatures are 
lower than the electroweak scale and numerically compute the amount of 
magnetic fields generated during inflation and reheating.
We find that, for lower reheating temperatures, the magnetic helicity 
decreases significantly. It is also possible to generate
fully helical magnetic fields by modifying the inflaton potential.
In both cases, the produced magnetic fields can be strong enough 
to explain the observed IGMFs, while avoiding the baryon 
isocurvature problem.
\end{abstract}
\maketitle

\section{I. Introduction}
\noindent
The existence of the cosmological magnetic fields is universal from small 
to the largest observed scales in the universe~\cite{2013A&ARv..21...62D}.
Of particular importance are the recent observational clues about 
intergalactic magnetic fields in the void 
regions~\cite{NeronovVovk10, Ackermann+18, MAGIC:2022piy}. 
Because of the difficulty for astrophysical mechanisms to account 
for them~\cite{Dolag:2010ni, Bondarenko:2021fnn}, 
the origin of intergalactic magnetic fields is often attributed to cosmological 
generation mechanisms in the early universe, which may, 
in combination with astrophysical amplification~\cite{Brandenburg:2004jv}, 
account for the origin of galactic and cluster magnetic fields.

Among the most popular scenarios is the inflationary magnetogenesis~\cite{turner1988inflation,ratra1992cosmological,garretson1992primordial}. 
If primordial magnetic fields are generated throughout the universe, 
they would naturally fill the void regions. 
However, various issues with inflationary scenarios have been pointed out, 
such as the backreaction~\cite{Bamba:2003av,Fujita:2015iga, Adshead:2016iae, Cuissa:2018oiw}, strong coupling~\cite{Demozzi:2009fu}, induced curvature perturbations~\cite{Barnaby:2012tk}, 
and Schwinger effect~\cite{Sobol:2018djj,Gorbar:2021rlt}. 
This research field has developed to date, with new techniques and scenarios 
being proposed to address them.

Recently, Ref.~\cite{Kamada:2020bmb} pointed out a new problem 
that is severe enough to invalidate most of the inflationary 
magnetogenesis models.
Any magnetogenesis scenario above the electroweak scale 
is strongly restricted because of the unavoidable contributions 
of magnetic fields to baryon isocurvature perturbations. 
We call it the baryon isocurvature problem, which will 
be explained below.
Their argument is so general that no effective countermeasures 
have been presented so far.
Therefore, it is crucial to clarify whether this problem 
is inevitable or not.

In this Letter, we show that low-scale inflationary scenarios 
are free from the baryon isocurvature problem and numerically 
demonstrate that helical coupling models~\cite{garretson1992primordial,Anber:2006xt} 
at low reheating temperatures can indeed account for 
the origin of cosmological magnetic fields.

Baryon number production from the magnetic fields occurs only above the electroweak scale. Therefore, if the universe never experiences a thermalized state 
with the temperature $T\gtrsim 100$ GeV, there would be no baryon isocurvature problem. Although such a scenario is contrary to the prevailing thermal history of the universe, it is still completely viable~\cite{Ferreira:2014hma, Fujita:2016qab, Kobayashi:2019uqs}.

Moreover, the analytic understanding of the magnetic field evolution has recently been updated based on the previously unrecognized conserved quantity of magneto-hydrodynamics, Hosking integral, and the wisdom in plasma physics, magnetic reconnection \cite{hosking2023cosmic,UCHIDA2023138002,uchida2024new}.
We use the new description for the evolution of 
cosmological magnetic fields and make a precise prediction of 
their present properties.

\section{II. Motivation}
\noindent
Let us first explain the motivation for considering a low-scale magnetogenesis scenario.
One may suppose that new physics beyond the electroweak scale generated the primordial magnetic fields.
Then, the magnetic fields are originally of the $\text{U}(1)_Y$ 
type rather than the $\text{U}(1)_\text{em}$ type.
During the electroweak symmetry breaking, as the gauge field configurations in the $\text{SU}(2)_L\times\text{U}(1)_Y$ sector gradually change, the $\text{U}(1)_Y$ magnetic helicity density is converted not only into the $\text{U}(1)_\text{em}$ magnetic helicity density but also into baryon number density, even taking into account the washout effect of the electroweak sphalerons, as a consequence of the chiral anomaly in the Standard Model~\cite{Kamada:2016cnb,Kamada:2016eeb}.

It is important to note that the produced baryon number densities 
have local fluctuations whose magnitude is proportional to 
$B^2\lambda(t_{\rm EW})$,
where $B$ and $\lambda$ 
are the typical strength and the coherence length of 
magnetic fields, respectively, and $t_{\rm EW}$ denotes 
the time of electroweak symmetry breaking~\cite{Kamada:2020bmb}. 
We have
\begin{align}\hspace{-2mm}
\overline{S^2_{B}}(t_{\rm BBN})\propto B^2 \lambda(t_{\rm EW})\;
\times(\text{neutron damping effect})\,,
\label{eq:FUbound}
\end{align}
where $\overline{S^2_{B}}(t_{\rm BBN})$ is the baryon 
isocurvature perturbation at big-bang nucleosynthesis, which is 
bounded from above, because otherwise, too much deuterium would be produced~\cite{Inomata:2018htm}.
We also have a damping factor, since the streaming of neutrons damps 
the baryon inhomogeneity on small scales, $\tilde{\lambda}\lesssim10^{-3}\,\text{pc}$ \cite{Inomata:2018htm,Kamada:2020bmb}.

\begin{figure}[t]
    \centering
    \includegraphics[keepaspectratio, scale=0.36]{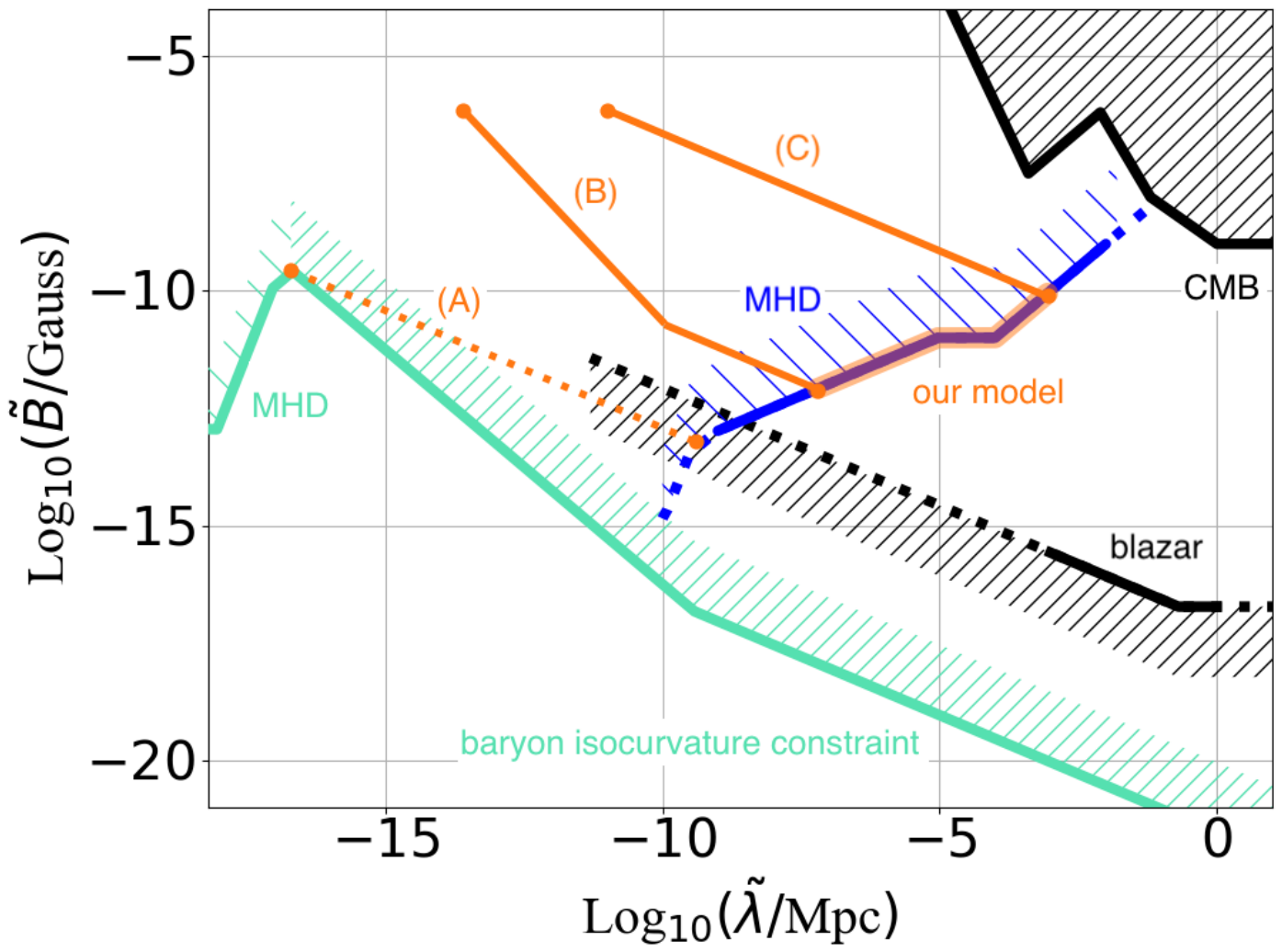}
    \caption{
    Constraints and predictions for cosmological magnetic fields.
    Apricot lines show the paths for representative evolution of three different scenarios: (A) the best scenario for magnetogenesis above the electroweak scale allowed by the upper bound from the baryon isocurvature (turquoise)~\cite{Inomata:2018htm} which does not fulfill the blazer lower bound (black hatched)~\cite{MAGIC:2022piy}, 
    (B) the GMSSM model with the reheating temperature $T_\text{reh}\sim100\;\text{GeV}$, 
    (C) the modified GMSSM model with $T_\text{reh}\sim1\;\text{MeV}$. 
    The break of line (B) corresponds to the change in the conservation law from $\tilde{B}^{4}\tilde{\lambda^5}\sim\text{const}$ to $\tilde{B}^{2}\tilde{\lambda}\sim\text{const}$ as the helicity fraction $\chi$ increases. 
    Our models predict the present-day magnetic field on the blue line (given by MHD analysis~\cite{hosking2023cosmic,UCHIDA2023138002,uchida2024new}) between (B) and (C), which 
    can successfully explain the intergalactic magnetic fields.
    } 
        \label{FU figure}	
\end{figure}

This constraint on $\overline{S^2_{B}}(t_{\rm BBN})$ translates into 
a tight bound on the cosmological magnetic fields at $t_{\rm BBN}$.
As shown in Fig.~\ref{FU figure}, this \textit{upper} bound is 
much \textit{lower} than the \textit{lower} bound on the present 
magnetic fields suggested by blazer observations.
Thus, any magnetogenesis scenario beyond the electroweak scale
cannot explain the origin of 
intergalactic magnetic fields in voids~\cite{Ackermann+18}, 
even if an optimistic evolution of fully helical magnetic fields is assumed.\footnote{Fully helical magnetic fields suffer from another problem, baryon overproduction \cite{Kamada:2016cnb}. To help the scenario, we need to assume anti-baryon generation mechanisms, while we need baryogenesis mechanisms to explain the baryon asymmetry of the universe for non-helical magnetic fields.
Fully helical fields are optimistic in the sense that their decay is milder than non-helical fields.} 
To evade the $\overline{S^2_{B}}$ constraint, 
we consider an alternative scenario, in which $\text{U}(1)_\text{em}$ magnetic fields are generated below the electroweak scale.

\section{III. Model}

\noindent
In this section, we describe our magnetogenesis scenario that is compatible with a low reheating temperature below the electroweak scale $\sim 100\,\text{GeV}$.
Let us consider the action ${\cal S}=\int {\rm d}^4x \sqrt{-g_M}\,\mathcal{L}$, 
where $g_M$ is the determinant of metric tensor $g_{\alpha\beta}$, and
\begin{align}\hspace{-1.5mm}
\mathcal{L}=-\frac{1}{2}\partial_\mu \phi \partial^\mu \phi -V(\phi) -\frac{1}{4}F_{\mu\nu}F^{\mu\nu}-\frac{1}{4}g\phi F_{\mu\nu}\tilde{F}^{\mu\nu},
\label{eq:axial coupling}
\end{align}
where $\phi$ is a pseudoscalar inflaton field with 
the potential $V(\phi)$, 
$F_{\mu\nu}\equiv \partial_\mu A_\nu-\partial_\nu A_\mu$ is 
the $\text{U}(1)_{\text{em}}$ gauge field strength, 
$\tilde{F}^{\mu\nu}\equiv 
\epsilon^{\mu\nu\rho\sigma}F_{\rho\sigma}/(2\sqrt{-g_M})$ 
is its dual with $\epsilon^{0123}=+1$, and $g$ is a coupling constant. This type of inflaton-gauge couplings was also studied in the context of magnetic field generations  \cite{Giovannini:1997gp,Anber:2006xt, 1992PhRvD..46.5346G}.
Assuming the spatially-flat Friedmann-Lema\^{i}tre-Robertson-Walker 
background, ${\rm d}s^2=-{\rm d}t^2+a^2(t) {\rm d}{\bm x}^2$,
the homogeneous part of the inflaton satisfies
\begin{equation}
\ddot{\phi}+3H\dot{\phi}+\partial_{\phi} V\simeq 0, 
\label{eq:t_EOM_phi}
\end{equation}
where $H=\dot{a}/a$ is the Hubble expansion rate, with a dot 
being the derivative with respect to the cosmic time $t$. 
On the right-hand side of Eq.~(\ref{eq:t_EOM_phi}), 
we ignored the backreaction from the gauge field, 
$g\langle\bm{E}\cdot\bm{B}\rangle$.
To ensure the validity of this approximation,  
we consider the coupling constant $g$ such that the energy density of 
magnetic fields ($\rho_{\rm B}$) is $10\%$ of the inflaton energy 
density ($\rho_{\phi}$) when it reaches the maximum value. 
We also make an optimistic assumption that the reheating completes 
at that time. In other words, we simply assumed that the $\phi$'s decay constant $\Gamma_\phi$ equals to $H(t)$ when the magnetic field takes its peak value. Hence we impose 
\begin{align}
\epsilon\equiv\frac{\rho_{\rm{B}}(t_\reh)}{\rho_\phi(t_\reh)}=\frac{\bm B^2}{\dot{\phi}^2+2V(\phi)}(t_\reh)=0.1\,, \label{eq:ratio_B}
\end{align}
where the subscript ``reh'' denotes the reheating completion. Note that the electromagnetic fields grow until $t_{\rm reh}$ and that they have similar 
energy densities $\rho_E\simeq \rho_B$ in our scenario, and hence the above condition is legitimate. Although the above discussion is for the backreaction 
of magnetic fields to the Friedmann equation, we have numerically confirmed that backreaction $g\langle\bm{E}\cdot\bm{B}\rangle$ to the inflaton equation of motion is also negligible.

\begin{figure}[t]
\centering
\includegraphics[keepaspectratio, scale=0.58]{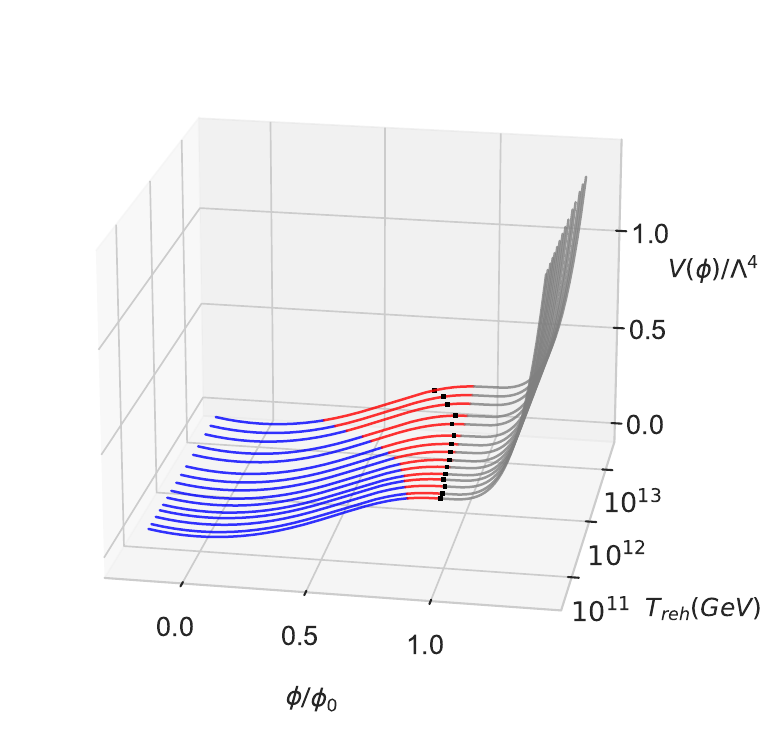}
\caption{The inflaton potential~\eqref{eq:GMSSMI} is divided into three color-coded regions; tachyonic instability (red), parametric resonance (blue), and 
no amplification of the electromagnetic fields (gray).
The border between the gray and red regions is determined by the criterion $g|\dot{\phi}|\geq2H$.
Each line represents a different choice of $\phi_0$, resulting in a different value of $T_\reh$. 
As $T_\reh$ decreases, the red region shrinks and tachyonic amplification 
tends to be less efficient. Black dots denote the end of inflation, $\dot{H}=-H^2$.
} 
\label{fig:3D}	
\end{figure}

We employ the inflaton potential in Generalized Minimal Supersymmetric Model (GMSSM)~\cite{martin2014encyclopaedia,lyth2007mssm},
\begin{align}
V(\phi)=\Lambda^4\qty[\frac{1}{2}\qty(\frac{\phi}{\phi_0})^2
-\frac{\alpha}{3}\qty(\frac{\phi}{\phi_0})^6
+\frac{\alpha}{10}\qty(\frac{\phi}{\phi_0})^{10}]\,, 
\label{eq:GMSSMI}
\end{align}
which respects the parity invariance under the 
sign change $\phi \to -\phi$.
The shape of the potential is shown in Fig.~\ref{fig:3D}.
This potential allows us to realize low reheating temperatures 
consistently with CMB observations. We fix $\Lambda$ and $\alpha$ to reproduce 
the observed amplitude and spectral tilt of curvature perturbations, respectively.

For the $\text{U}(1)_{\text{em}}$ gauge field, we choose the 
radiation gauge, $A_0=0$ and $\nabla\cdot\bm{A}=0$, and quantize it as
\begin{align}\hspace{-1.5mm}
A_i(t, \bm{x})
=& \sum_{\lambda=\pm} \int \frac{{\rm d}^3 k}{(2\pi)^3} e^{i \bm{k \cdot x}}
\notag\\
&\times e_{i}^{(\lambda)}(\hat{\bm{k}}) \left[\hat b_{\bm{k}}^{(\lambda)} \mcA_\lambda(k,t) +\hat b_{-\bm{k}}^{(\lambda) \dag} \mcA_\lambda^*(k,t) \right],\label{eq:quantization}
\end{align}
where $\bm{e}^{(\pm)}(\hat{\bm{k}}) $ is a circular-polarization 
vector with the comoving wavenumber ${\bm k}$,  
$\hat b_{\bm{k}}^{(\lambda)\dag}$ and $\hat b_{\bm{k}}^{(\lambda)}$ 
are standard creation and annihilation operators, respectively, 
and $\mcA_\lambda(k,t)$ is a mode function satisfying
\begin{equation}
\ddot{\mcA}_{\pm} +H \dot{\mcA_\pm} +
\frac{k}{a}\left(\frac{k}{a} \mp g  \dot{\phi} \right)
\mcA_\pm=0. \label{eq:t_EOM_A}
\end{equation}
In this model, the electromagnetic fields experience two distinct stages of 
the amplification~\cite{Fujita:2015iga, Adshead:2016iae, Cuissa:2018oiw}:
(i) When the inflaton velocity is large enough to satisfy the inequality $g|\dot{\phi}|\gtrsim 2H$, either of two circular-polarization modes 
$\mcA_\pm$ effectively obtains a negative mass squared ({\it i.e.}, 
the parenthesis in Eq.~\eqref{eq:t_EOM_A} becomes negative) and grows due to tachyonic instability\footnote{In terms of the conformal 
time $\eta=\int\,a^{-1}{\rm d}t$, the mode 
functions $A_{\pm}$ takes the form 
$\partial_{\eta}^2 {\cal A}_{\pm}+k^2 
[1 \pm 2\xi/(k \eta)]{\cal A}_{\pm}=0$, 
where $\xi=g \dot{\phi}/(2H)$. 
In the limit $k \eta \to 0$ with a constant $\xi$, we have the following 
asymptotic solution
$A_{+} \to e^{\pi |\xi|}/\sqrt{4\pi k |\xi|}$ 
and hence $A_{+}$ is significantly amplified 
for $\xi>{\cal O}(1)$.},
(ii) While the inflaton oscillates after inflation, parametric resonance 
takes place~\cite{finelli2001resonant} and both polarization modes are equally amplified.
We emphasize that the helicity of the magnetic field ({\it i.e.}, a significant imbalance between $\mcA_+$ and $\mcA_-$) is produced only in the first stage.

Once $\mcA_{\pm}(k,t)$'s are computed, the comoving magnetic field 
strength and its coherence 
length are evaluated, respectively, as
\begin{align}
\tilde B^2(t)&\equiv a^{4}B^2 = \int \frac{\dd^3 k}{(2\pi)^3}\, k^2 \Big(
|\mcA_+|^2+|\mcA_-|^2\Big)\,,
\label{eq:mag integral}\\
\tilde \lambda(t) &\equiv \frac{\lambda}{a} 
= \frac{\int \dd^3 k\ (2\pi/k)\, k^2 \left(
|\mcA_+|^2+|\mcA_-|^2\right)}{\int \dd^3 k\ k^2 \left(
|\mcA_+|^2+|\mcA_-|^2\right)}\,,
\label{eq:integral}
\end{align}
where the scale factor is normalized to be  $a=1$ today.
We also pay particular attention to the helicity fraction,
\begin{align}
\chi(t)\equiv\frac{\int \frac{\dd^3 k}{(2\pi)^3}\, k \Big( |\mcA_-|^2 - |\mcA_+|^2\Big)}{\int \frac{\dd^3 k}{(2\pi)^3}\, k \Big( |\mcA_-|^2 + |\mcA_+|^2\Big)}\,,
\label{eq:chi}
\end{align}
where $\chi$ indicates how helical the magnetic fields are,
and $\chi=\pm 1$ corresponds to maximally helical magnetic fields. 

For simplicity, we assume that the inflaton instantaneously decays 
into radiation at $H_\reh$. Then, we can approximate the reheating temperature as
$T_\reh \simeq \qty[90\Mpl^2 H_\reh^2/(100\pi^2)]^{1/4}$~\cite{kolb2003reheating}, 
with the scale factor $a_\reh\simeq 5.5\times 10^{-32}\sqrt{\Mpl/H_\reh}$, 
where $\Mpl$ is the reduced Planck mass. 
We ignore the effect of charged particles before the reheating completion.

\section{IV. Numerical calculation}

\noindent
We numerically solve Eqs.~\eqref{eq:t_EOM_phi} and \eqref{eq:t_EOM_A} 
combined with the Friedmann equation, $3\Mpl^2 H^2 \simeq \rho_\phi$,
until the magnetic field generation weakens and the energy fraction 
starts decreasing.
The background inflaton reaches the inflationary attractor 
before the horizon-crossing of the CMB modes. Note that the electromagnetic 
fields are initially in the Bunch-Davies vacuum.
We repeatedly run simulations by varying the model parameters 
corresponding to different reheating temperatures $T_{\rm reh}$. In Appendix A, we give the parameter sets used in our calculations and 
describe our parameter selection strategy.

Let us mention here whether it is necessary to 
take the inflaton inhomogeneity 
into account. The inflaton perturbation 
may be enhanced if the inflaton-gauge coupling is sufficiently large. However, for an increasing coupling constant, it is known that the backreaction of gauge fields at the background level first becomes important, 
before the inflaton inhomogeneity grows significantly \cite{Fujita:2015iga}. 
Since we are considering 
a regime where the backreaction from gauge fields is negligible, we do not implement 
the effect of the inflaton inhomogeneity 
on the generation of magnetic fields.

\begin{figure}[t]
\centering
\includegraphics[keepaspectratio, scale=0.57]{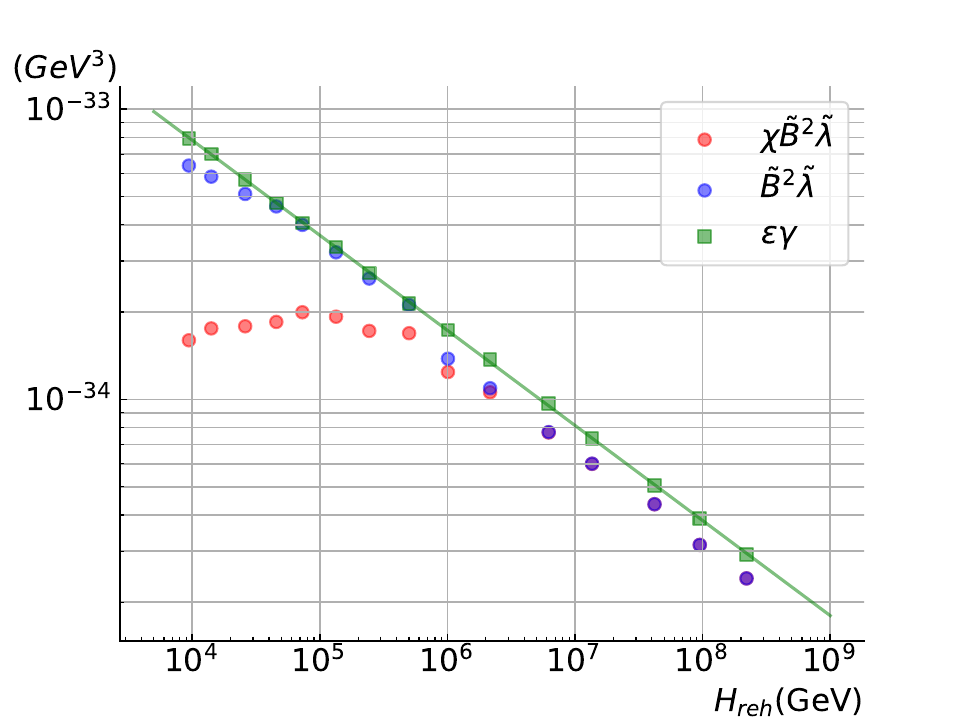}
\caption{The numerical results for $\tilde B^2_\reh \tilde \lambda_\reh$ (blue) 
and $\chi_\reh \tilde B^2_\reh \tilde \lambda_\reh $ (red) against $H_\reh$. 
An analytic estimate for $\tilde B^2_\reh \tilde \lambda_\reh$ (green) derived 
below Eq.~\eqref{eq:h_max} agrees well with the former. 
For $H_\reh \lesssim 10^6~\GeV$, the helicity fraction $\chi_\reh$ 
significantly decreases from unity as $H_\reh$ decreases. } 
\label{fig:h_max}	
\end{figure}

We are interested in $\tilde B^2 \tilde \lambda $ to which 
the blazer observations are sensitive. In Fig.~\ref{fig:h_max}, 
we present our numerical result of $\tilde B^2\tilde \lambda (t_\reh)$ 
versus the Hubble parameter $H_\reh$ at the temperature $T_{\rm reh}$. 
We also plot $\chi \tilde B^2\tilde \lambda (t_\reh)$ and 
an analytic estimate of $\tilde B^2\tilde \lambda(t_\reh)$, 
which is derived as follows.
The inflaton equation of motion is given by 
$\partial_\mu\qty[a^3\partial^\mu\phi-(g/2)\epsilon^{\mu\nu\rho\sigma}
A_\nu\partial_\rho A_\sigma]=a^3 \partial_\phi V$. 
Note that the second term on the left-hand side was 
neglected in Eq.~\eqref{eq:t_EOM_phi}, whose
approximation assumes that $|a^3\dot{\phi}| \gg 
|(g/2) \epsilon^{0\nu\rho\sigma}A_\nu\partial_\rho A_\sigma|$.
The homogeneous oscillating inflaton 
at $t_\reh$ is approximately given by
$|\dot{\phi}|\simeq\sqrt{3}\Mpl H_\reh$.
As a large-volume average, we have
$|\epsilon^{0\nu\rho\sigma}A_\nu\partial_\rho A_\sigma|
\simeq \chi \tilde B^2\tilde \lambda/2\pi$.
However, if the magnetic fields are non-helical ($|\chi|\ll 1$), 
it would underestimate its local value. 
Local fluctuations yield $|\epsilon^{0\nu\rho\sigma}A_\nu\partial_\rho A_\sigma| 
\simeq \sqrt{\langle\left(\epsilon^{0\nu\rho\sigma}A_\nu\partial_\rho A_\sigma\right)^2\rangle}\simeq  \tilde B^2 \tilde \lambda/2\pi$, 
which could dominate the homogeneous value. 
Thus, we adopt the latter and obtain a generic upper bound as
\begin{align}
\tilde{B}^2_\reh \tilde{\lambda}_\reh \lesssim \gamma,\quad
\gamma \equiv \frac{4\sqrt{3}\pi}{g}a_{\reh}^3\Mpl H_\reh. \label{eq:h_max}
\end{align}
As shown in Eq.~\eqref{eq:ratio_B}, the maximum backreaction of 
the magnetic field to the Friedmann equation is of order $\epsilon \rho_\phi$. 
We assume that the magnetic-field backreaction to the inflaton equation 
of motion is at most of order $\epsilon$ times the scalar-field contributions, 
which is consistent with our numerical results.
Then, the actual values of $\tilde B^2_\reh \tilde \lambda_\reh$
are expected to be of the order of $\epsilon \gamma$. 
In Fig.~\ref{fig:h_max}, we see an excellent agreement 
between this analytical estimate (green) and the numerical results (blue).

\begin{figure}[t]
\centering
\includegraphics[keepaspectratio, scale=0.57]{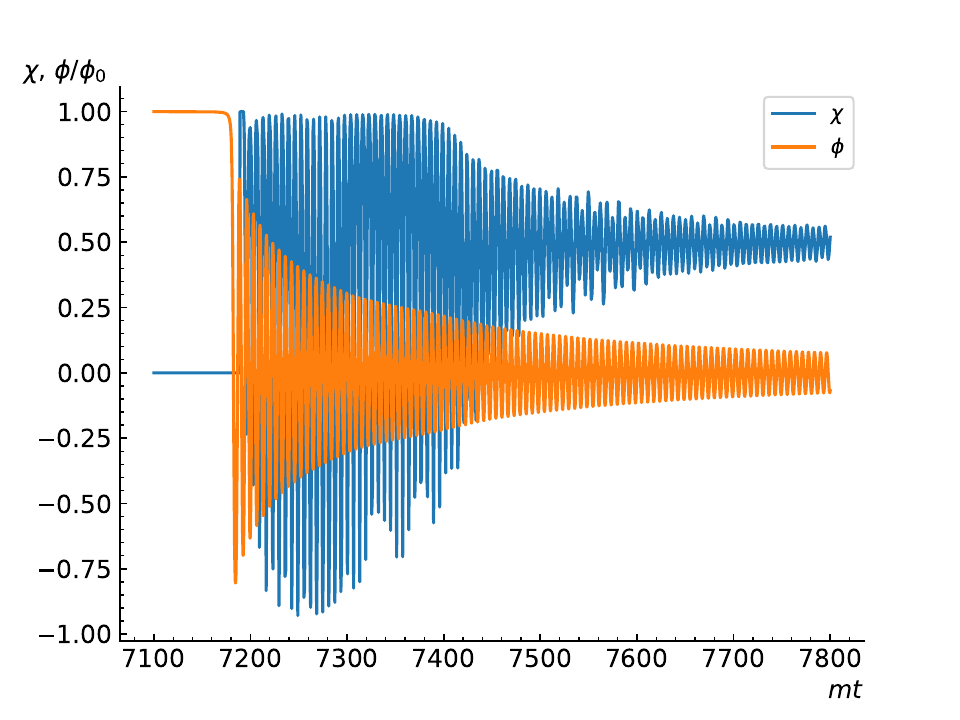}
\caption{The time evolution of the homogeneous inflaton $\phi$ (orange) and 
the helicity fraction $\chi$ (blue) in an intermediate case of $T_\reh\simeq 10^{11}\GeV$. 
In such a case, tachyonic instability generates only $\mathcal{O}(1)$ imbalance of 
two circular-polarization modes ({\it e.g.}, $|\mcA_-|= 2|\mcA_+|$) until 
the inflaton oscillation begins. For each half oscillation, 
parametric resonance alternately amplifies $\mcA_-$ and $\mcA_+$ 
by up to $\mathcal{O}(10)$ and $\chi$ flips its sign. 
As the inflaton amplitude decreases, parametric resonance 
becomes less effective and $\chi$ converges to an $\mathcal{O}(1)$ value. 
For lower (higher) $T_\reh$, the effect of tachyonic instability 
is less (more) significant, which leads to $\chi\ll1\  (\chi\simeq 1)$.
}
\label{fig:chi0.03}
\end{figure}

As seen in Fig.~\ref{fig:h_max}, 
the produced magnetic fields are significantly non-helical ($\chi\ll 1$)
for $H_\reh \ll 10^6~\GeV$. This is because 
the amplification during the first stage due to 
tachyonic instability is no longer efficient and
the magnetic fields are mostly generated by parametric resonance 
in the second stage for low reheating temperatures. 
In Fig.~\ref{fig:3D}, we plot the region of the potential in 
which tachyonic instability occurs before the end of inflation, 
namely $g|\dot{\phi}| \geq 2H$.
Such a region shrinks as $T_\reh$ decreases because inflation ends 
with the rapid growth of $|\dot{\phi}|$, and then the inflaton 
immediately starts oscillating, as is typical in small-field models.
In Fig.~\ref{fig:chi0.03}, we show the numerical evolution for 
$\phi(t)$ and $\chi(t)$ in an intermediate case where $\chi$ 
converges to $\simeq 0.5$. 
The helicity fraction $\chi$ rapidly oscillates 
in phase with the inflaton, which indicates that the magnetic fields are generated primarily by parametric resonance. Nevertheless, the final value of $\chi$ is determined by the imbalance between the two circular-polarization modes produced by relatively inefficient tachyonic instability, since parametric resonance equally amplifies both of the modes.

In much lower $T_\reh$ cases, it is numerically challenging to track the extreme 
slow-roll evolution and obtain fine-tuned values of $\alpha$ as pointed out 
in \cite{martin2014encyclopaedia}.
To extrapolate them to lower $T_{\rm reh}$, we find fitting functions 
of the comoving magnetic-field amplitude, comoving coherence length, 
and helicity fraction, as (see Appendix B for details) 
\begin{align}
\tilde{B}_\reh&\simeq 7.0\times10^{-7}~\text{G}\,, 
\label{eq:strength_B}
\\
\tilde{\lambda}_\reh&\simeq2.7\times10^{-14}~\Mpc
\qty(\frac{T_\reh}{100~\GeV})^{-0.68}\,, 
\label{eq:lambda_fit}\\
\chi_\reh&\simeq3.1\times10^{-6}\qty(\frac{T_\reh}{100~\GeV})^{0.55}\,, 
\label{eq:chi_fit}
\end{align}
for $T_\reh \ll 10^{12}~\GeV$. For the fitting of $\chi_\reh$, we used only the numerical results 
of the parameteric-resonance dominant regime ($H_\reh\le 10^6~\GeV$), 
while we exploited all the data points for the $\tilde \lambda_\reh$ fitting.
$\tilde{B}_\reh$ does not depend on $T_{\rm reh}$, because 
$B_\reh \propto \sqrt{2\epsilon\rho_{\phi,\reh}} \propto H_\reh$
and $a_\reh\propto H_{\reh}^{-1/2}$ are cancelled out.
As $T_\reh$ decreases, we are left with a smaller residual difference 
in the almost equally produced two circular-polarization modes of 
the magnetic fields, which leads to
$\chi_\reh\propto T_\reh^{0.55}$ in Eq.~\eqref{eq:chi_fit}.

Before turning to our main results, we consider maximizing the produced 
magnetic fields by slightly modifying the inflaton potential.
To maintain the tachyonic instability region of the potential even 
for very low $T_\reh$, we attach a quadratic function to 
the potential~\eqref{eq:GMSSMI} as
\begin{align}
V_{\rm mod}(\phi)=
\begin{cases}
V(\phi-\phi_1) & ({\rm for}~\phi-\phi_1\geq\phi_f) \\
m_0^2\phi^2/2 & ({\rm for}~\phi-\phi_1<\phi_f)
\end{cases}, 
\label{eq:potential2}
\end{align}
where $m_0 \equiv \Lambda^2/(100\phi_0)$, $\phi_f$ denotes $\phi$ 
at the end of inflation, and $\phi_1$ is the appropriate shift for 
the two parts of the potential to be smoothly connected at $\phi_f$. 
$\phi_1$ is uniquely determined by matching the potential at $\phi_f$.
Right after inflation ends, the inflaton gently rolls down on 
$m_0^2\phi^2/2$ where tachyonic instability efficiently occurs 
irrespective of $T_\reh$. 
The shrinkage of the red region seen in Fig.~\ref{fig:3D} 
does not occur in this case.
We numerically show that our scenario with this new potential always produces 
fully helical magnetic fields with $\chi=1$.
$\tilde B_\reh$ is the same as Eq.~\eqref{eq:strength_B}, 
and we fit (see Appendix B for details)
\begin{align}
\tilde{\lambda}_\reh = 1.4\times10^{-14}~\Mpc
\qty(\frac{T_\reh}{100\GeV})^{-0.57}. 
\label{eq:lambda_fit2}
\end{align}
Although the modified potential~\eqref{eq:potential2} is somewhat artificial, 
the implications of examining it are twofold: first, to validate our understanding 
of why $\chi_\reh$ reduces as $T_\reh$ decreases in the original case and second, 
to provide an example of single-field inflation models generating much stronger 
magnetic fields than the previous case.

\section{V. Magnetic fields at present time}
\noindent
Summarizing the results so far, either of Eqs.~\eqref{eq:strength_B}, \eqref{eq:lambda_fit}, and \eqref{eq:chi_fit} or Eqs.~\eqref{eq:strength_B}, \eqref{eq:lambda_fit2}, and $\chi_\reh=1$ set 
the initial conditions for the subsequent evolution of magnetic fields. 
Now, let us discuss the implication of these results in light of observational 
constraints on the cosmological magnetic field in today's universe.

In tracking the evolution until the recombination epoch, 
we can use the magnetic helicity conservation due to the large electric conductivity \cite{2013A&ARv..21...62D},
\begin{align}
\chi \tilde{B}^{2}\tilde{\lambda}
=\text{const.}\qquad (t_\reh\le t\le t_\rec)\,.
\label{eq:MagneticHelicityConservation}
\end{align}
The MHD analysis has provided another condition 
at the recombination time 
$t_\rec$~\cite{hosking2023cosmic,uchida2024new},
\begin{align}
\left(\dfrac{\tilde{B}_\text{rec}}{10^{-12}\;\text{G}}
\right)^2\left(\dfrac{\tilde{\lambda}_\text{rec}}{10^{-8}\;
\text{Mpc}}\right)^{-1}\sim1\,,
\label{eq:ResultantMF4}
\end{align}
which we adopt as a universal approximation, valid for $10^{-13}\,{\rm G}\lesssim\tilde{B}_\text{rec}\lesssim10^{-10}\,{\rm G}$, of the blue line in Fig.~\ref{FU figure}.
Since maximally helical magnetic fields keep $|\chi|=1$ 
during the evolution, the above two equations suffice 
to determine $\tilde{B}_\rec$ and 
$\tilde{\lambda}_\text{rec}$ independently
for the modified potential (\ref{eq:potential2}) with $\chi_\reh=1$.

If $\chi_\reh<1$, however, we need yet another equation 
to find $B_\rec, \lambda_\rec$ 
and $\chi_\rec$.
Recently, it has been pointed out that the Hosking integral is conserved for non-helical magnetic fields~\cite{2021PhRvX..11d1005H,2022JPlPh..88f9002Z,2023Atmos..14..932B},
\begin{align}
\tilde{B}^4_\reh \tilde{\lambda}^5_\reh=\text{const.}
\qquad (t_\reh\le t\le t_\rec,\,|\chi|\ll 1).
\label{eq:HoskingIntegralConservation}
\end{align}
For maximally helical magnetic fields, the Hosking integral 
is ill-defined and hence this constraint is irrelevant. 
Although it is not well-established how partially helical 
magnetic fields (e.g., $|\chi|=\mathcal{O}(0.1)$) behave, 
we assume that Eq.~\eqref{eq:HoskingIntegralConservation} holds~\cite{uchida2024new}.

Combining Eqs.~\eqref{eq:MagneticHelicityConservation}, \eqref{eq:ResultantMF4} and \eqref{eq:HoskingIntegralConservation}, we obtain
\begin{align}
    &\left(\tilde{B}_\text{rec}/10^{-12}\;\text{G},\  \tilde{\lambda}_\text{rec}/10^{-8}\;\text{Mpc}\right)\notag\\
    &\sim\left\{
    \begin{array}{cc}
       \left(1\times \mathcal{T}_\reh^{-0.03},\, 2\times \mathcal{T}_\reh^{-0.07}\right)  &  (T_\reh \lesssim 570 {\rm MeV})\\[4pt]
       \left(0.5\times \mathcal{T}_\reh^{-0.24},\, 0.2\times \mathcal{T}_\reh^{-0.49}\right)  &  (T_\reh\gtrsim 570 {\rm MeV})\\[4pt]
       \left(30\times \mathcal{T}_\reh^{-0.14},\, 800\times \mathcal{T}_\reh^{-0.29}\right) & ({\rm modified\ model})
    \end{array}
    \right.,
    \label{eq:ResultantMF5}
\end{align}
with $\mathcal{T}_\reh\equiv T_\text{reh}/100\,\text{GeV}<1$.
Their evolution paths and the resultant magnetic fields are shown in Fig.~\ref{FU figure}.
Note that,  conservation laws for non-helical fields, Eqs.~\eqref{eq:MagneticHelicityConservation} and \eqref{eq:HoskingIntegralConservation}, imply $\chi_{\rm rec}=(\tilde{B}_{\rm rec}/\tilde{B}_{\rm reh})^{-6/5}\chi_{\rm reh}$. By substituting our model prediction for the GMSSM model, the assumption of $\chi_{\rm rec}\ll1$ breaks down if  $T_\text{reh}\gtrsim 570\;\text{MeV}$, implying that
the magnetic fields become maximally helical by the recombination. 
In such cases, 
we use only Eqs.~\eqref{eq:MagneticHelicityConservation} 
and \eqref{eq:ResultantMF4} 
with $\chi_\rec=1$.

We expect that the intergalactic magnetic fields in the void remain frozen in the comoving sense after the recombination epoch.
Thus we identify their present strength and coherence length with the comoving ones at recombination, $B_0=\tilde{B}_\rec$ and $\lambda_0=\tilde{\lambda}_\rec$.
Now, our results are compared to the latest 
observational constraint~\cite{MAGIC:2022piy},
\begin{align}
\hspace{-2mm}
\left(\hspace{-0.5mm}\dfrac{B_0}{2\times10^{-17}\;\text{G}}\hspace{-0.5mm}\right)^{\hspace{-1mm}2}
\left(\hspace{-0.5mm}\dfrac{\lambda_0}{0.2\;\text{Mpc}}\hspace{-0.5mm}\right)
\hspace{-0.5mm}
\gtrsim\hspace{-0.5mm} 1~~\;\text{for}\,\; \lambda_0\hspace{-0.5mm}\ll\hspace{-0.5mm}0.2\;\text{Mpc}.
\label{eq:ObservationalBound}
\end{align}
As seen in Fig.~\ref{FU figure}, our scenario satisfies 
this constraint for any $T_\reh<100~{\rm GeV}$ and thus 
successfully explains the origin of cosmological magnetic fields.

\section{VI. Conclusion}
\noindent
In this paper, we have proposed an inflationary magnetogenesis 
scenario that is consistent with current observations. 
Our scenario can explain the origin of the void magnetic field 
without producing baryon isocurvature perturbations. 
The key point is that primordial magnetogenesis occurs below 
the electroweak scale, which makes the baryon 
isocurvature problem irrelevant. 
We have performed numerical simulations for low-energy inflation potentials, 
and successfully produced magnetic fields that exceed the blazar constraint.

Although our results capture the universal properties of low-energy magnetogenesis 
scenarios, calculations for other kinds of potentials, especially more natural ones that realize $\chi_\reh=1$, would be interesting. Furthermore, by taking into account the backreaction~\cite{Bamba:2003av,Fujita:2015iga, Adshead:2016iae, Cuissa:2018oiw}, Schwinger effect~\cite{Sobol:2018djj,Gorbar:2021rlt}, and the electric conductivity 
of the universe~\cite{Bassett:2000aw,Fujita:2019pmi}, more precise results will 
be obtained in the numerical simulation. We leave these issues for future works.

\vspace{5mm}

\acknowledgments
\section{Acknowledgement}
We thank Kohei Kamada for useful discussions.
The work is supported by JSPS KAKENHI Grant Nos.~JP23KJ0642 (F.U.), JP23K03424, JP20H05854 (T.F.), JP22K03642 (S.T.) and also by the Forefront Physics and Mathematics Program to Drive Transformation (F.U.) and Waseda University Special Research Project No.~2023C-473 (S.T.).

\appendix
\section{Appendix A: Parameters}
Here we describe the way to choose parameters for 
the numerical calculations. 
The inflaton potential must be consistent with the observed data of 
CMB temperature anisptropies. 
Once we fix $\phi_0$ in Eq.~\eqref{eq:GMSSMI}, 
$\alpha$ is almost uniquely determined by requiring 
the agreement with the observed spectral index $n_s$ of scalar perturbations. 
It must be in the range of $n_s=0.966\pm0.004$ \cite{2020A&A...641A...6P}, 
and $\alpha$ needs to be severely fine-tuned as mentioned previously. 
Also, the value of $\Lambda$ is determined by the CMB constraint on 
the amplitude of curvature perturbations. The upper bound on the tensor-to-scalar ratio is trivially satisfied due to a small slow-roll parameter during inflation.

The remaining parameter is the Chern-Simons coupling constant, $g$. 
To fix this value, we define $t_\reh$ as the time at which each magnetic field strength reaches its maximum, and we repeat the calculations for 
different coupling constants $g$. 
Then, we find the appropriate value of $g$ satisfying 
the condition $\epsilon(t_\reh)=0.1$. 
The parameters we used for the GMSSM potential (\ref{eq:GMSSMI}) 
are shown in Tab.~I. The ones for the modified potential 
(\ref{eq:potential2}) are also given in Tab.~I\hspace{-1.2pt}I.

\begin{table}[h]
\centering
\begin{tabular}{ccl}
\hline
 $\phi_0/\Mpl$ &  ~~~$g\Mpl$~~~ & $1-\alpha$ \\
\hline
    0.013 &   105.2 &   $1.0\times10^{-13}$ \\
    0.016 &    97.30 &   $2.0\times10^{-13}$ \\
    0.020 &    87.60 &   $5.0\times10^{-13}$ \\
    0.025 &    79.90 &   $1.1\times10^{-12}$ \\
    0.030 &    74.00 &   $2.4\times10^{-12}$ \\
    0.040 &    65.90 &   $7.0\times10^{-12}$ \\
    0.050 &    60.20 &   $1.9\times10^{-11}$ \\
    0.070 &    53.50 &   $6.5\times10^{-11}$ \\
    0.100 &    46.40 &   $2.8\times10^{-10}$ \\
    0.140 &    40.18 &   $9.8\times10^{-10}$ \\
    0.220 &    33.55 &   $6.0\times10^{-9}$ \\
    0.300 &    29.90 &   $2.0\times10^{-8}$ \\
    0.500 &    24.70 &   $1.4\times10^{-7}$ \\
    0.700 &    21.40 &   $6.0\times10^{-7}$ \\
    1.000 &    18.65 &   $2.3\times10^{-6}$ \\
\hline
\end{tabular}
\caption{The parameters used for the numerical calculations 
of the GMSSM potential (\ref{eq:GMSSMI}).}
\end{table}

\begin{table}[h]
\centering
\begin{tabular}{ccl}
\hline
 $\phi_0/\Mpl$ &  ~~~$g\Mpl$~~~ & $1-\alpha$ \\
\hline
    0.013 &    29.90 &   $1.0\times10^{-13}$ \\
    0.016 &    27.53 &   $2.0\times10^{-13}$ \\
    0.020 &    25.38 &   $5.0\times10^{-13}$ \\
    0.025 &    23.35 &   $1.3\times10^{-12}$ \\
    0.030 &    21.45 &   $2.4\times10^{-12}$ \\
    0.040 &    18.74 &   $7.0\times10^{-12}$ \\
    0.050 &    17.36 &   $1.9\times10^{-11}$ \\
    0.070 &    16.85 &   $6.4\times10^{-11}$ \\
    0.100 &    16.50 &   $2.8\times10^{-10}$ \\
    0.140 &    16.25 &   $9.8\times10^{-10}$ \\
\hline
\end{tabular}
\caption{The parameters used for 
the modified potential (\ref{eq:potential2}).}
\end{table}

\section{Appendix B: The extrapolated results}
\begin{figure}[H]
\centering
\includegraphics[keepaspectratio, scale=0.4]{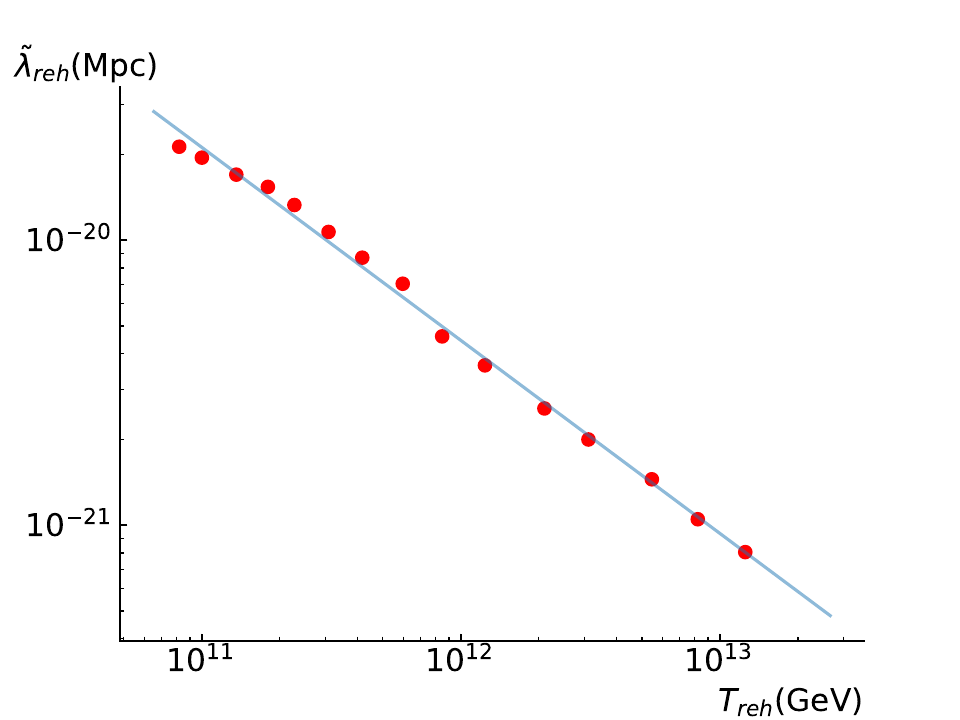}
\caption{The relation between the comoving coherence length 
$\tilde{\lambda}_\reh$ and reheating temperature $T_\reh$. 
The blue solid line shows the fitting function given by Eq.~\eqref{eq:lambda_fit}. 
}
\label{fig:T_lambda}	
\end{figure}

\begin{figure}[H]
\centering
\includegraphics[keepaspectratio, scale=0.4]{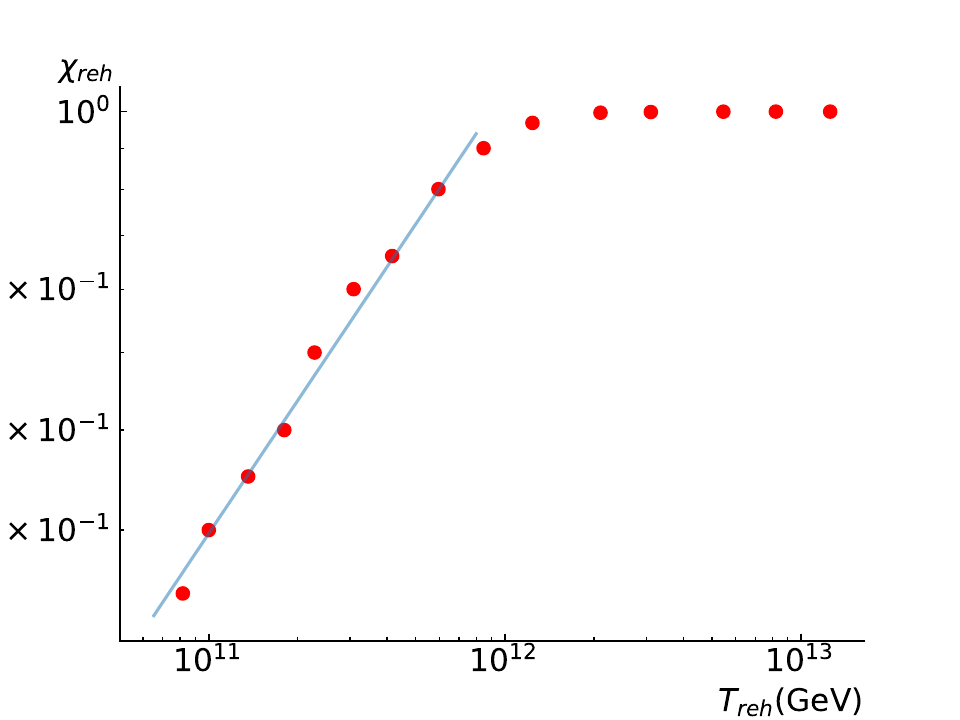}
\caption{The relation between the helicity fraction $\chi_\reh$ 
and $T_\reh$. The helicity fraction saturates at unity for 
$T_\reh \gtrsim 10^{12}$~GeV. 
For lower reheating temperatures, $\chi_\reh$ 
exhibits the power-law behavior.
}
\label{fig:T_chi}	
\end{figure}

To obtain the values of the comoving coherence length $\tilde{\lambda}_\reh$ and the helicity fraction $\chi_\reh$ for lower $T_\reh$, we extrapolate the numerical results by assuming that they are power-law functions of $T_\reh$. 
The fitting functions given by Eqs.~\eqref{eq:lambda_fit} and \eqref{eq:chi_fit} 
are shown in Figs.~\ref{fig:T_lambda} and \ref{fig:T_chi}, respectively. 
While we use all the points in Fig.~\ref{fig:T_lambda} for fitting 
$\tilde \lambda_\reh$, we only use 9 data points on the low $T_\reh$ 
side for fitting $\chi_\reh$ in Fig.~\ref{fig:T_chi}, 
which follows a power-law relation.

\begin{figure}[H]
\centering
\includegraphics[keepaspectratio, scale=0.4]{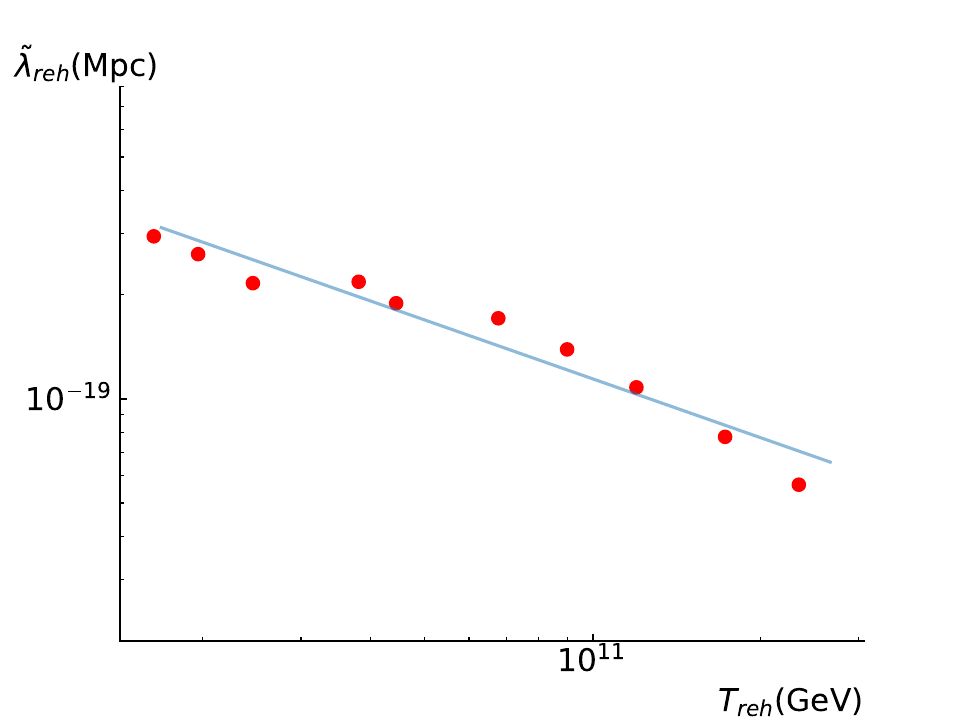}
\caption{The relation between the comoving coherence length $\tilde{\lambda}_\reh$ and reheating temperature $T_\reh$ 
for the potential \eqref{eq:potential2}. 
}
\label{fig:T_lambda2}	
\end{figure}

Similarly, we extrapolate $\tilde{\lambda}_\reh$ for the 
modified potential \eqref{eq:potential2}
and obtain the fitting formula \eqref{eq:lambda_fit2}
shown in Fig.~\ref{fig:T_lambda2}.
Note that $\chi_\reh$ is always unity and 
we do not need to extrapolate it in this model.

\bibliographystyle{unsrt}
\bibliography{arXiv}
\end{document}